\documentclass[12pt,letterpaper]{article}

\newcommand{\ubar}{\ensuremath{\overline{u}} }
\newcommand{\dbar}{\ensuremath{\overline{d}} }
\newcommand{\sbar}{\ensuremath{\overline{s}} }

\newcommand{\lf}{\leaders \hbox to 1em {\hss .\hss}\hfill}
\begin{document}
\pagestyle{empty}
\null
\vskip 2 true in

\centerline{\bf \Large Key Issues in Hadronic Physics}
\vskip 1 true in

\begin{abstract}
A group of fifty physicists met in Duck, NC, Nov. 6-9 to discuss
the current status and future goals of hadronic physics. The
main purpose of the meeting was to
define the field by identifying its key issues, challenges, and
opportunities.
The conclusions, incorporating considerable input from
the community at large, are presented in this white paper. 
\footnote{This document is to be presented at the Town Meeting
at Jefferson Lab Dec. 1-4 and made available to NSAC to aid
in the long range planning process. It does not represent
a response to the NSAC charge, a historical review of hadronic
physics, or an endorsement of any particular experimental effort.
RHIC physics is being reviewed in a separate process and is therefore
not discussed herein.}
\end{abstract}

\vfill\eject
\null
\vskip 1 true in
\begin{center}
{\bf Contents}
\end{center}
\vskip .5 true in
{{\bf 1 Introduction} \lf 1}

\noindent
{{\bf 2 Fundamental Problems in Strong Interaction Physics} \lf 3}

{2.1 Parameters of QCD \lf 3}

{2.2 How Does QCD Work? \lf 3}

{2.3 QCD in the Wider World \lf 6}

\noindent
{{\bf 3 The Quark and Gluon Structure of Hadronic Matter as Probed through Hard Scattering and Form Factors} \lf 8}

{3.1 Hadronic Structure \lf 8}

{3.2 Hadronization: the Dynamics of Physical State Formation \lf 12}

{3.3 The Role of Quarks and Gluons in Nuclei and Partonic Matter Under Extreme Conditions \lf 13}

\noindent
{{\bf 4 Spectroscopy} \lf 14}

{4.1 Mesons \lf 15}

{4.2 Baryons \lf 18}

\noindent
{{\bf 5 The Chiral Structure of Matter, Form Factors, and Few Body Nuclei} \lf 20}

{5.1 The Chiral Structure of Matter \lf 20}

{5.2 Nucleon Electromagnetic Form Factors \lf 21}

{5.3 Few Body Nuclei \lf 22} 

\noindent
{{\bf 6 Models of the Quark Structure of Matter} \lf 23}

{6.1 Structure \lf 23}

{6.2 Substructure \lf 24}

{6.3 Partonic Region \lf 25}

\noindent
{{\bf 7 Tools} \lf 26}

{7.1 Experiment \lf 26}

{7.2 Theory \lf 27}

\noindent
{{\bf Contributors} \lf 27}

\vfill\eject
\pagestyle{plain}
\pagenumbering{arabic}

\section{ Introduction}

Strong interaction physics poses a wealth of fundamental questions with
profound significance for our understanding of Nature and the structure of
the matter of which we and our universe are composed. Answering these
questions lies at the heart of contemporary nuclear science and will
have deep impact on particle physics, astrophysics, and cosmology.

The field of hadronic physics is the study of strongly interacting matter in all its
manifestations and the understanding of its properties and interactions in
terms of the underlying fundamental theory, Quantum Chromodynamics (QCD).
It is a vibrant and growing field, which now encompasses a large fraction
of nuclear physics and has attracted a significant number of particle
physicists. The field has a long history, starting with phenomenological
descriptions of hadron-hadron interactions and the hadron spectrum and
continuing to present day ideas on the quark-gluon structure of hadrons,
heavy quark symmetry, effective field theory, the quark-gluon plasma, and
novel color superconducting phases of matter among a host of others.
Although many of its deepest questions have challenged us for decades, we
now have within our grasp unprecedented opportunities for fundamental
progress. Recent advances in computational technology, lattice field
theory algorithms, continuum model building, accelerator beam quality, and
detector design have led us to the threshold of developing a true
understanding of the fundamental mechanisms of QCD and the ability to
solve nonperturbative QCD quantitatively.  This report describes the
present status of hadronic physics, the scientific opportunities it
provides, and the means by which the national hadronic physics community
is poised to exploit these opportunities.

{\bf The primary goals of hadronic physics are to determine the relevant
degrees of freedom that govern hadronic phenomena at all scales, to
establish the connection of these degrees of freedom to the parameters and
fundamental fields of QCD, and to use our understanding of QCD to
quantitatively describe a wide array of hadronic phenomena, ranging from
terrestrial nuclear physics to the behavior of matter in the early
universe. }

The theoretical foundations and extensive experimental tests of the
standard model in general and QCD in particular are so compelling that the
focus is not on {\it testing} QCD but rather on {\it understanding} QCD.
Specific objectives of the field, which are addressed in more detail in
the main text of the report, include the following:

\begin{itemize}
\itemsep=0pt

\item{\bf Determine the parameters of QCD.}
 
The fundamental scale, $\Lambda_{QCD}$, which sets the scale for all
strong interaction phenomena, the masses of quarks which ultimately
control details of hadron spectroscopy, and the QCD vacuum $\Theta$
parameter controlling the violation of CP symmetry need to be determined
precisely.

\item{\bf Understand the origin and dynamics of confinement.}

The remarkable fact that the fundamental constituents of composite
hadrons, quark and gluons, cannot be removed from hadrons and examined in
isolation sets hadrons apart from all other known composite systems.
Whereas lattice calculations clearly indicate the formation of tubes of
gluonic fields connecting colored charges, we need to understand from
first principles why flux tubes are formed, how they relate to 
the confinement of color charge, and 
the role that they play in the structure and dynamics of hadrons.
Experimental exploration of the full spectrum of states composed of quarks
and gluons will be an important tool in attaining this understanding.

\item{\bf Understand the origin and dynamics of chiral symmetry breaking.}

The spontaneous breaking of chiral symmetry, responsible for the existence
of light pions, their dynamics, and the masses of hadrons needs to be
understood directly in terms of QCD. We need to understand the physical
origin, topological or otherwise, of the quark zero modes generating the
chiral condensate, and to understand the relationship between the
deconfinement and chiral phase transitions at finite temperature.

\item{\bf Understand the quark and gluon structure of hadrons based on
QCD.}

One of the principal Science objectives in the Department of Energy
Strategic Plan is to develop a quantitative understanding of how quarks
and gluons provide the binding and spin of the nucleon based on QCD. This
objective is a central focus of our field.

\item{\bf Understand the relation between parton degrees of freedom in the
infinite momentum frame and the structure of hadrons in the rest frame.}

Deeply inelastic scattering experiments, a major quantitative tool for
exploring the quark and gluon structure of hadrons,
measure correlation functions along the light cone and thus naturally
determine probability distributions of partons in the infinite momentum
frame. We need to develop physical insight and quantitative tools to 
relate parton distributions to the structure of hadrons in their rest
frame.

\item{\bf Develop quantitatively reliable models and approximations to
QCD.}

The understanding and synthesis of a wealth of existing and forthcoming experimental 
data requires the 
development of reliable models. This process will be aided by qualitative 
insights and constraints arising from the development of controlled expansions of QCD such as 
the heavy-quark, large-$N_c$, and chiral limits; from the techniques of 
effective field theory; and from quantitative and qualitative lattice 
results.


\item{\bf Explore the role of quarks and gluons in nuclei and matter under
extreme conditions.}

{}From the modification of the quark-gluon structure of a nucleon when it is
immersed in the nuclear medium within a nucleus to the novel phases and
behavior of matter in neutron stars, supernovae, or the early universe,
there are a host of fundamental questions that hinge crucially on
developing the ability to understand and quantitatively solve QCD.

\end{itemize}

The body of this report is organized as follows. We begin by discussing in
more detail the fundamental problems arising in strong interaction
physics.   The next two sections describe two ways to gather experimental
information on hadronic physics: using deeply inelastic scattering to study partons 
in hadrons,  and studying quarks and gluons in the excited state spectrum of
mesons and baryons.
The role of models is discussed in section 5.
Finally,
new theoretical and experimental tools that promise unprecedented
opportunities for fundamental progress in hadronic physics are highlighted
in section 6.

\newpage

\section{Fundamental Problems in Strong Interaction Physics}

To place the subsequent details of experimental and theoretical
exploration of hadronic physics in context, it is useful to begin by
considering the truly fundamental problems arising in contemporary
hadronic physics.

\subsection {Parameters of QCD}

There is compelling evidence that in additional to its beauty and
theoretical appeal, the QCD Lagrangian completely describes the strong
interactions. Hence the challenge is to determine its parameters, solve
it, and understand it.

The fundamental scale, $\Lambda_{QCD}$, or equivalently the running
coupling constant $\alpha_S$, emerges from QCD through the phenomenon of
dimensional transmutation so it is crucial to determine it accurately. At
present, the numerical solution of lattice QCD provides one of the most
precise values of $\Lambda_{QCD}$, which is also in good agreement with
state-of-the art experimental determinations. With requisite effort, this
evaluation can be improved by an order of magnitude, thereby providing an
essential parameter needed to understand the unification of the
fundamental forces.

Experiments on the electric dipole moment of the neutron indicate that
the value of the $\theta$ angle, another fundamental parameter of QCD,
is very small. This leads to a major puzzle called the strong CP 
problem. Since one possible resolution would be for the up quark mass, 
$m_u$, to be zero, it is particularly important to measure the 
renormalization group invariant mass ratio, $\frac{m_d-m_u}{m_d+m_u}$. 
A combination of theoretical analysis based on chiral perturbation 
theory and numerical lattice calculations make it possible to calculate 
this ratio convincingly within the next five years, and this is a high
priority, showcase  calculation. It is also of interest and feasible to
determine the absolute masses of the strange quark,  m$_s$(m$_Z$) and of  the 
heavy quarks.

\subsection{How does QCD work?}

Although a quarter of a century has passed since the experimental
discovery of quarks in the nucleon and the invention of QCD,
understanding how QCD works remains one of the great puzzles
in many-body physics. One major challenge arises from the fact that
the degrees of freedom  observed in low energy phenomenology are totally
different from those  appearing in the QCD Lagrangian. Indeed, unlike
any other many-body  system, the individual quark and gluon
constituents making  up a proton cannot even be removed from the system
and examined  in isolation. In addition, in the past, there
were no quantitative tools to calculate non-perturbative QCD. Now,
however, the combination of theoretical tools and experimental probes
presently available offers an unprecedented opportunity to make decisive
progress in understanding how QCD works.

\subsubsection{Fundamental Aspects} 

There are three fundamental questions upon
which all else hinges. What are the degrees of freedom and  mechanisms
responsible for confinement, for chiral symmetry breaking, and for U(1)
symmetry breaking? Understanding these mechanisms from first principles
and developing the tools to calculate them quantitatively will provide the
foundation for understanding hadronic physics. 

Several analytical approaches provide valuable insight and theoretical
guidance. Semiclassical objects including instantons, monopoles, and
vortices identify essential nonperturbative effects that may play
significant roles in confinement, chiral symmetry breaking, and $\eta'$
mass generation. The strong coupling expansion and the related
emergence of flux tubes provides strong insight into the physics of confinement.
Expansions around three complementary analytically tractable limits
provide valuable insight into the physical regime. The heavy quark limit
emphasizes the universal adiabatic behavior of glue and light quarks in
the presence of static color sources. The chiral limit emphasizes the
role of pion degrees of freedom in the kinematical regime in which
excitations of heavier degrees of freedom are suppressed. 
Studies based on this limit can describe the long range part of 
hadronic structure and interactions in a controlled way. Finally,
the large N$_c$ limit emphasizes the simplifications in the classes of 
diagrams that contribute, and the mean field effects that arise, when the
number of colors is large.

The advances in lattice field theory and the availability of very large
scale computers make it possible for the first time to complement these
analytic approaches with definitive numerical solutions of QCD
for a large class of important problems. In addition to enabling
quantitative calculation of physical quantities like the chiral
condensate, topological susceptibility, string tension, and interface
energy between confined and deconfined phases, the lattice provides
important opportunities for insight. For example, one can 
directly explore the dependence of these quantities on the number of flavors, the number of
colors, and the values of quark masses and thereby test theoretical
mechanisms in ways that are impossible with laboratory
experiments. In addition, one can directly determine the configurations which
dominate the QCD path integral and attempt to extract qualitative features of 
them. Finally, one can use lattice calculations to constrain and improve models,
for example by evaluating overlaps between 
exact wavefunctions and model {\it Ans\"atze}.

\subsubsection{Hadron Structure -- Two Complementary Perspectives}

It is natural to view the structure of hadrons from two very different
and complementary perspectives. The challenge is not only to complete our
understanding from each viewpoint, but also to relate the degrees of
freedom arising in one description to those
appropriate to the other.

{\it Snapshots in quarks and glue}

Asymptotic freedom enables us to use the tools of perturbative QCD to
precisely characterize deeply inelastic lepton scattering from hadrons.
Since these experiments measure correlations along the light cone, the
resulting structure functions  are naturally described by the
light cone distributions of quarks and gluons, or equivalently,
quark and gluon distributions in the infinite momentum frame. These
experiments first revealed quarks and gluons in the nucleon and have now
determined the light cone quark distribution, helicity
distribution, and gluon distribution in great detail. An important
conceptual advantage of these distributions is that the quarks and gluons
they measure are directly related to the quark and gluon degrees of
freedom appearing in the QCD Lagrangian. One limitation is that
they tell us the probability of finding a quark
with a given momentum fraction, $x$, but  yield no information about the
phase of the amplitude.

Whereas perturbative QCD is crucial in extracting these
distributions from experiments, it is totally inadequate for the deeper
challenge of calculating them from first principles. Thus, it is a major
development that contemporary theory has become sufficiently powerful
to calculate low moments of these distributions nonperturbatively.

Experiments now have the potential to tell us in detail
how the total spin of the proton is divided between the spin
and orbital angular momentum of quarks and gluons. Measurements of
higher twist effects can specify, for example, correlations between
quarks and gluons.  A particularly interesting and novel possibility
arises from the fact that the
gluon component of the nucleon grows as the momentum fraction $x$
decreases, so that very low $x$  physics provides a new regime dominated by a
sea of gluons. Thus, experimental study of very low $x$ offers the
tantalizing possibility of exploring this new gluon dominated regime in
which essential simplifications in QCD may occur and a new form of
universal behavior may arise.

{\it Pictures with dressed quarks}

The other natural perspective from which to view hadron structure is in
the rest frame; which is appropriate for consideration of spectroscopy and
measurements of quantities like the charge radius, magnetic moment, and
axial charge. This is the frame in which the familiar quark model
works far better than we can presently justify from first principles.
Here, the degrees of freedom are not simply those of the Lagrangian and
we need to understand their microscopic foundations. What is the relevant
quark degree of freedom, the so-called constituent or dressed quark, and
how is it related through its cloud of gluons and quark-antiquark pairs
to the quark and gluon fields of the underlying Lagrangian? How do the
resulting quasiparticles interact? Although we
know that essentially all of the mass and half the momentum and spin of the
nucleon are carried by glue, what is the role of this glue in the nucleon,
and how can we observe it? Is it concentrated in flux tubes associated
with confinement, and if so, can we find unambiguous signatures in
spectroscopy such as states with exotic quantum numbers arising from
excitation of flux tubes?  To what extent are lumps of glue associated
with instantons responsible for the nonperturbative interactions between
quarks in light hadrons? How does the dressed quark picture manifest the 
underlying chiral symmetry structure of QCD and thereby produce 
the known long distance behavior of hadrons described by chiral
perturbation theory?  Well chosen spectroscopy, in concert with 
theoretical analyses 
of hadron wave functions and vacuum structure, offers the potential for crisp
answers to many of these difficult conceptual questions  that have
puzzled us for decades.  

\subsubsection{Nuclear Physics}

The next intellectual challenge is to go beyond the physics 
of a single hadron and understand essential aspects of nuclear physics 
from first principles.  In thinking about many-nucleon systems, one
immediately faces the question of the origin of the nuclear energy
scale.  Why, when the natural energy scale of QCD is of the order of
hundreds of MeV, is the nuclear binding energy per particle so small, of
the order of 10 MeV? Does it arise from complicated details of near
cancellations of strongly attractive and repulsive terms in the nuclear
interaction or is there some deeper reason for this scale to arise?

{\it Calculations in small $A$ nuclei from QCD}

The large separation between the hadronic energy scale 
and the nuclear binding scale renders it difficult to apply QCD directly
to understand the physics of small $A$ nuclei. However, quantitative 
calculations based on effective field theory techniques that arise
from chiral symmetry provide an alternative approach. Traditionally,
this method has been applied to the physics of pions in the 
context of chiral perturbation theory. Currently, it is being 
extended to address many nucleon interactions. When combined 
with first principles calculations of the low energy constants 
from QCD, these effective field theories may have the potential to
provide a systematic and quantitative tool to study low energy properties of light nuclei. 
If quantitative models of low
energy QCD are developed, it would be valuable to use them to derive the gross
features of low-A nuclear physics and thereby illuminate how the
macroscopic features of nuclear physics emerge from the underlying quark
and gluon degrees of freedom.

{\it Insight into large $A$ from QCD}

It is harder to envision understanding the physics of large $A$ nuclei,
nuclear matter, and neutron star matter from effective field theory alone.
Given the success of nuclear many-body theory based on phenomenological
potential fits to nucleon-nucleon phase shifts, it would be valuable to
understand qualitative features of these potentials directly from
QCD -- for example the origin of the hard core and the spin and
isospin dependence of the nucleon-nucleon interaction.  The heavy quark limit is particularly valuable in
this regard, since one can calculate the adiabatic potential between
hadrons containing one heavy quark on the lattice and thereby explore the
role of light quark exchange and gluon exchange in detail.
The large $N_c$ limit is also useful in elucidating certain
features of hadron-hadron interactions.

In addition to understanding the structure of nuclei {\it per se}, it is
also of interest to understand the behavior of nucleons within nuclei.
Over a decade ago, deeply inelastic scattering experiments by the  EMC
collaboration and its successors showed that the quark distribution
in a nucleon immersed in the nuclear medium differs substantially from
that in free space. Whereas the calculational tools at that time were
inadequate to discriminate between several plausible mechanisms, we now
have the opportunity to clarify this physics.

\subsection{QCD in the wider world}

QCD is the essential ingredient of the Standard Model 
that is not yet under quantitative control. Precise 
calculations based on it are necessary to understand 
a variety of phenomena in high energy physics and astrophysics 
that extend far beyond the traditional boundaries of nuclear 
physics. Furthermore, there are illuminating
connections between QCD and condensed matter 
physics from which both fields can benefit.

\subsubsection{High Energy Physics and Astrophysics}

The design and interpretation of experiments to search for
fundamental physics beyond the standard model rely on
firm quantitative control of QCD. At present, the largest
uncertainty in many high energy physics calculations
comes from incalculable strong interaction matrix elements. For example,
the matrix element 
$\langle P| m_s \bar{s}s | P\rangle$ is necessary to
determine whether the neutralino, a supersymmetric particle,
can help solve the ``dark matter'' puzzle. Calculation of this 
strangeness matrix element in the proton state is necessary 
to quantify the coupling of the neutralino to matter.

Other calculations are necessary to understand the relation between
the fundamental quark and gluon interaction parameters with the
mixing of $K$ and $\bar{K}$ mesons or with the electric dipole 
moment of the neutron.
These calculations can shed light
on how one of the fundamental symmetries of nature, 
a combination of charge conjugation and parity, 
is violated. 

The mass difference between the proton and the neutron,
$m_N-m_P$, is an energy scale that is crucial to
the structure of our world. A grand challenge 
that would truly test our mastery of QCD would be to calculate it from
first principles. This involves an interesting interplay of
electromagnetic  interactions and the difference between  the up and down quark 
masses.

\subsubsection{Extreme Conditions in the Lab and the Cosmos} 

QCD is essential to answer questions related to the
physics of the early universe and high energy astrophysics.
For example, just after the 
big bang, when matter was extremely hot, QCD predicts 
that quark and gluon degrees of freedom dominated the world. 
As the universe cooled, these degrees of freedom were bound into
hadrons, reducing the number of degrees of freedom 
dramatically. The temperature and the nature of this 
qualitative change in the phase of matter is an extremely 
interesting question for cosmologists. The search for the 
new state of matter at high temperatures and densities
is currently being undertaken in relativistic
heavy ion experiments.

Observation of ultra high energy cosmic rays with energies of the order 
of $10^{11}$ GeV implies that neutrinos with similar energies 
must also be present. If the large arrays of neutrino detectors
that are planned in the future can detect this flux of 
neutrinos, they can act as laboratories for a new form of
deeply inelastic scattering experiments. Learning about the 
structure functions of the proton at low $x$ will be crucial to interpret 
the results of these experiments.

The physics of QCD at high densities plays a critical
role in determining the physics of neutron stars and 
supernovae. The equation of state at finite density is essential for
quantitative calculations of the astrophysics of neutron stars. The
excitation spectrum of hadronic matter must be understood to  predict
their neutrino emission spectra.   Recently the theoretical exploration of novel
phenomena in quark matter at  high density, such as color
superconductivity and color-flavor locking, has given new impetus to
understanding QCD at high density and its astrophysical consequences. As
our quantitative understanding of this regime develops, it may also
provide new insight into the domain of ordinary nuclear densities.
Indeed, one of the fundamental questions we must ultimately address is
the nature of the true ground state of hadronic matter. Is it really
true, as is usually assumed, that the ground state of hadronic matter
resembles a collection of conventional nuclei, or is such a state a metastable 
excitation of the true ground state comprised of up, down, and strange
quarks? The definitive calculation  of the ground state of matter is another
worthy Grand Challenge for hadronic physics.

\subsubsection{Connections with Condensed Matter Physics}

The phenomena and challenges that arise in hadronic physics have much in
common with those arising in strongly interacting condensed matter
systems. For example,
the complex  dynamics of QCD simplifies at 
certain critical points in the QCD phase diagram leading to universal
critical behavior,  which can be modeled with much simpler degrees of
freedom such as those arising in the Ising model.  Similarly, the
superfluid phases of $^3$He have much in common with phases in
dense QCD. Lattice QCD can be formulated in the language of
quantum spin systems, to which cluster algorithms and the insights from
dimensional reduction directly apply. Finally, the notorious fermion sign
problem that pervades Monte Carlo calculations in condensed matter
problems also arises in QCD, and the invention of techniques to solve it
in QCD offers corresponding benefit in condensed matter physics. 
Thus, the deep interconnections between the physics of strongly
interacting systems with many degrees of freedom in condensed matter
physics and hadronic physics offers the potential for  mutually
beneficial sharing of insights and techniques. 


\section{The Quark and Gluon Structure of Hadronic Matter as
Probed through Hard Scattering and Form Factors}


Nucleons are the primary building 
blocks of atomic
nuclei and other hadronic matter in the universe. The first direct evidence that
the nucleon is a composite particle came from the experimental measurement 
of elastic form factors in the 1950's. The quark substructure
of the nucleon was clearly revealed through electron-proton deeply inelastic
scattering (DIS) at SLAC. Following these pioneering
discoveries, a great amount of information about the partonic (quark 
and gluon) structure of hadronic matter has been learned through 
measurements of form factors and quark and gluon distributions. However, 
our knowledge is still far from complete. Some crucial 
questions in this field remain open:

{\it
\begin{enumerate}
\item What is the {\rm structure} of hadrons in terms of their 
       quark and gluon constituents?
\item How do quarks and gluons \textit{evolve} into hadrons via the dynamics
	of confinement?  
\item What is the role of quarks and gluons in the structure of atomic
	{\rm nuclei}? How can nuclei be used to study matter under
	{\rm extreme conditions}?
\end{enumerate} 
}
The answer to these questions is the missing key to our ultimate
understanding of the microscopic structure of matter. In the following
subsections, we examine the physical content of these questions and
explore future opportunities in this field.

\subsection{Hadronic Structure}
Understanding the structure of the nucleon in terms of the quark and
gluon constituents of QCD is one of the outstanding fundamental problems
in physics.  The field-theoretical nature of strong interactions 
leads to the picture of a nucleon as an  ensemble of a 
large and ever-changing number of constituents.  A major aim of 
experiments through the next decade is to take detailed ``snapshots" 
of this structure at various levels of resolution.  The highest 
resolution is provided by highly energetic
projectiles, which interact with individual quarks, antiquarks, and
gluons inside a proton or neutron.  These interactions, being sensitive
to the motion of the struck particle, can map the probability for
finding the various constituents as a function of $x$ -- the fraction 
they carry of the nucleon's overall momentum.  Such detailed maps 
will provide a crucial test of QCD-based calculations of nucleon structure.
Indeed, a number of basic features have yet to be delineated or understood.
At the same time,
less energetic projectiles must be used to obtain a lower resolution,
but more global, view of the nucleon's properties, {\it e.g.} elastic form factors,
which describe the overall distribution of charge, magnetism, and the magnetic
dipole moment of baryonic resonances.

{}From a large body of available experimental data, 
the up and down quark distributions at moderate $x$
are found to be consistent with the simple picture of quark 
models. Gluons also play a crucial role since 
they carry nearly 50\% of the nucleon's momentum. 
Over the past 5 years, precision measurements from polarized DIS 
indicate that the quark spins account for only about 30\% of 
the nucleon's spin, in marked contrast with the constituent quark picture,
where the quark spins carry all of the nucleon's spin.
The credibility of the data is backed 
by the verification of the so-called Bjorken sum rule -- a relation which
follows directly from QCD. In addition, contrary to naive 
expectations, data from Drell-Yan and electroproduction experiments 
show a pronounced excess of \dbar over $\ubar$ quarks 
at intermediate values of $x$, possibly indicating the importance
of Goldstone boson degrees of freedom in nucleon structure. 

Unfortunately we do not have a detailed or comprehensive knowledge of 
nucleon structure.
Exciting and fundamental discoveries have yet to 
be made in multiple frontiers, as illustrated by the following
examples. 
 

%
\subsubsection{Strangeness in Nucleons}

Strange quarks in the nucleon arise from
``vacuum fluctuations".
However, the pattern of strange quark
effects shows interesting irregularity.  
Some observables have little or no influence from strange quarks, while
strangeness makes a significant contribution to others.
{}From the study of pion-nucleon scattering, it was found
that strange quarks are responsible for a sizable fraction 
of the nucleon mass. Furthermore, polarized DIS
data with the assumption of SU(3) quark flavor symmetry 
hint that $s$ quarks may carry as much as $-$10\% of the nucleon 
spin. On the other hand, data from DIS with neutrino 
beams and other experiments indicate that $s$ quarks account for only a few percent of 
the nucleon's total momentum and that the $s$ and $\sbar$ 
distributions seem similar. Some of the important
theoretical and experimental questions to be answered  
include: Are $s$ and $\sbar$ distributions really similar? If
so, why? How do strange quarks contribute to the nucleon's
magnetic and electric form factors? How can we understand
the pattern of the strange quark effects? Experiments, which are actively under way,
address some of these questions, such as the strange form factors.

%
\subsubsection{Spin of the Nucleon}

How does the proton get its spin? Polarized DIS data have
shown that quark spins account for only about 30\% of the proton spin.
Where are the missing contributions? Besides the 
quark orbital angular momentum, gluons are expected to be strongly 
polarized. Indeed a precision QCD analysis of polarized
DIS data and recent measurements of hadron-pair production have given 
a preliminary indication of a large gluon polarization $\Delta G$. 
In the near future, determining $\Delta G$ with good 
precision is one of the most important objectives in high-energy
spin physics. While proton-proton collision experiments are expected to
play a crucial role in understanding the behavior of  polarized gluons, 
high-energy polarized electron-proton collisions can provide 
interesting complementary information.

One possible source of the strong deviation of the measured quark
spin contribution from that expected by assuming that it is all carried
by valence quarks is a significant polarization of sea quarks. 
A direct measurement of this will
provide an immediate test of various, more sophisticated, nucleon models,
which give qualitatively different predictions for the polarization
of the antiquarks. Experiments in progress 
and planned will directly 
study this question. 

\subsubsection{Structure Functions at Large $x$}

Our knowledge of the quark distributions at large $x$ 
is sketchy at best. The regime $x \rightarrow 1$ represents a 
fascinating kinematic limit, where a single parton is 
responsible for the entire momentum of the proton. 
The main problem with existing data on the ratio $u(x)/d(x)$ 
is that the experiments rely on the use of the deuteron to 
provide a neutron target. The Fermi motion and binding 
of the neutron in this nuclear bound state introduce large 
uncertainties in the partonic interpretation of the
data in the limit $x \rightarrow 1$. New experiments that 
eliminate this problem are a high priority. In the same
limit, it is expected that the struck parton carries the entire 
spin of the proton as well as its momentum, and so the double spin 
asymmetry $A_1$ should approach unity. This expectation must be tested
soon. The importance of parton distributions
at large $x$ is also reflected in their use as
essential input to high energy experimental searches for physics
beyond the Standard Model.

A promising new tool for studying large-$x$ behavior is found in 
the use of quark-hadron duality, first discovered by Bloom 
and Gilman. Recent precise measurements suggest that 
the nucleon resonance region
can be used to determine reliably the large $x$ behavior of structure
functions, which would be difficult to measure using the canonical
kinematics of DIS. 
Duality in semi-inclusive processes remains to be explored.
%

\subsubsection{New Parton Distributions}

Besides the unpolarized $q(x)$ and helicity-dependent $\Delta q(x)$
quark distributions, 
a complete description of nucleon structure at leading order requires 
the \textit{transversity} distribution $\delta q(x)$. This
distribution describes quark polarization within a transversely 
polarized nucleon and does not mix with gluons under scale evolution. 
In the absence of relativistic effects, the 
transversity distribution $\delta q(x)$ should be equal to 
$\Delta q(x)$, and this provides a ``baseline'' for 
our understanding of this, as yet unmeasured, distribution. 
The first moment of $\delta q(x)$ (termed the tensor charge of the nucleon)
offers a promising point for comparison with theory.
Because $\delta q(x)$ decouples from inclusive DIS, semi-inclusive
experiments with transversely polarized targets
are needed for dedicated measurements of this unknown quantity.

A significant development in hadronic physics over the last several years 
is the identification of a new class of parton distributions, known as
\textit{Generalized Parton Distributions} (GPD). 
Probed primarily in exclusive measurements, the GPDs describe 
hard scattering processes that involve the \textit{correlations} 
between partons. This new formalism offers an exciting bridge between 
elastic and deeply inelastic scattering: in different kinematic limits of 
the GPDs, one recovers the familiar elastic form factors and DIS 
structure functions of the proton. Clearly, a mature description of 
the partonic substructure of the nucleon, beyond the naive picture
of collinear non-interacting quarks, must involve a description of these
partonic correlations. 
Further, GPDs have a direct connection to the unknown parton orbital angular
momentum (which is an essential contribution to the total spin of the nucleon) 
and to the impact parameter dependence of parton distributions.
Experimentally, exclusive scattering measurements at large $Q^2$
and small $t$, the so-called deep-exclusive scattering (DES), 
are just beginning. 
It is essential to continue vigorous theoretical and experimental 
studies of these interesting new parton distributions.

High $t$ exclusive reactions are the most direct way of observing partonic
correlations.  The ability to carry out such experiments
has been demonstrated in JLab experiments on elastic form factors, 
$N\to N^*$ amplitudes, wide angle Compton scattering,
and $\phi$ photoproduction.
At a specific value of $t$, these different reactions 
probe different characteristics of the GPDs.  Thus, these 
quantities put very precise  constraints of any models of generalized parton 
distributions.
Since $t$ is directly related to the mean transverse momentum  
of the participating partons, the $t$ dependence of these reactions 
yields a measure
of the transverse high momentum components of the parton distributions
and correlations. 


%

\subsubsection{The Partonic Substructure of Mesons and Hyperons, SU(3) Flavor Symmetry}

Whereas the proton and neutron are the ``building blocks''
of atomic nuclei, pions and kaons (and mesons in general) supply the 
``mortar'' that holds the nucleus together. At a fundamental level, 
pions and kaons are the 
Goldstone bosons of spontaneously broken chiral symmetry. 
A familiar example of Goldstone bosons comes from the existence 
of phonons in crystalline materials due to spontaneously
broken translational symmetry.

We know little about the partonic substructure of mesons.
Since these particles are unstable on time scales of order $10^{-8}$ sec
or less, they cannot be used as viable fixed targets. 
However, some measurements have been made using either meson beams or by
scattering from
the virtual meson cloud around the nucleon.
These first data are exciting, but of low precision. A 
new experimental program is required if one wants to answer these questions:
Is the structure of mesons similar to that of baryons? Do sea quarks and 
gluons play as prominent a role in the substructure of the chiral Goldstone
boson as they do in the proton? And most fundamentally, how is 
the transition from partonic degrees of freedom to Goldstone modes 
accomplished?

If one were to map out the substructure of the pion and kaon, 
important tests of so-called SU(3) flavor symmetry might be performed. 
One of the basic precepts of the strong interaction is that it is 
``flavor blind'': only the quark mass term of the QCD Lagrangian
distinguishes one quark flavor from another. 
Since the light quark masses are small compared with the physical 
scale of strong interactions, the structure of the meson should
have an approximate flavor symmetry. 
Independent measurements of the substructure of several
of the pseudo-scalar mesons would provide a powerful test of this 
fundamental precept.

Furthermore, experimental techniques exist that enable the measurement
of the partonic substructure of \textit{hyperons}. These are  
 $J^P$ = $\frac{1}{2}^+$ 
baryons like the proton that contain a strange quark in the valence
sector. Hyperons are also being studied with models and lattice QCD.
These investigations will permit the detailed exploration 
of SU(3) symmetry in 
the baryon sector, where extensive information on the two 
lightest members is already available. 
Lastly, DES is capable of 
comparing parton densities in different baryons: nucleons, 
$\Delta$-isobars, and hyperons and to probe  short distance
$q\bar q$ wavefunctions of different mesons.

\subsection{Hadronization: The Dynamics of Physical State Formation}

A fundamental question in hadronic physics is 
how a quark or gluon from high-energy scattering 
evolves into a hadron. This process is known as
\textit{hadronization}, and is a clear manifestation
of color confinement: the asymptotic physical states 
detected in experiment must be color-neutral hadrons.
Hadronization also appears in an astrophysical context, as part of 
the transition from a deconfined state of free quarks and gluons in
the Big Bang into stable protons, which provide the seeds for
nuclear synthesis. Understanding fragmentation
in spin-dependent processes, the use of fragmentation as a tool for hadron 
structure study, and probing the global structure of 
the hadronic final state are likely to be the main 
themes of future investigation in this area.

\subsubsection{Testing the Dynamics of Confinement} 

Hadronization is a complex, non-perturbative process that
is related to both the structure of hadronic matter and 
to the long-range dynamics of confinement. Understanding 
hadronization from  
first principles has proven very difficult. However, over the last 
two decades, progress has been made in phenomenological 
descriptions of hadronization, such as the Lund model.
One immediate goal is 
to extend and test the consequences of the model in different
physical domains. For instance, how well can the model describe the 
data at lower center-of-mass energy, where jet formation does not
occur?
More interestingly, how should spin degrees of 
freedom be incorporated in fragmentation processes? The
latter is particularly important because spin admits a rich 
variety of fragmentation functions, posing 
challenges to any fragmentation model. A
fundamental question is how, and to what extent, the spin
of a quark is transfered to its hadronic daughters. 
A related goal is to understand the quantum state of the $q\bar q$
pair that emerges from breaking the color flux tube. 


%
\subsubsection{A Tool for Hadron Structure Studies}

In the nuclear physics laboratory, hadronization has emerged over the 
last 5 years as a \textit{tool} of profound importance in the analysis 
of hadronic structure functions: a new generation of experiments is 
exploiting the fact that semi-inclusive DIS measurements may, 
through fragmentation functions, ``tag''
particular flavors of struck quarks.  New varieties of semi-inclusive
and exclusive processes have also introduced new \textit{classes} 
of hadronization observables. For example, the measurement of 
the transversity distribution $\delta q(x)$ relies on the 
participation of the T-odd fragmentation function
$H_1^\perp(z)$ with attendant \textit{phase coherence} in the 
final state.  With better understanding of the 
spin transfer mechanism, useful information could be
gleaned about the spin structure of the produced 
hadron itself, such as the $\Lambda$ baryon whose spin
can be measured from the angular distribution of its decay products.

%
%


\subsection{The Role of Quarks and Gluons in Nuclei, and Partonic Matter
Under Extreme Conditions}

Most of the observable matter in the universe is 
contained in the form of atomic nuclei. The interaction between 
protons and neutrons is responsible for \textit{nuclear binding} and 
may be described with good success using effective theories where 
exchanged mesons (predominantly pions) serve as mediators. How is this
binding effect manifested in the underlying quark and gluon 
degrees of freedom? How important is the effect of the
nuclear modification of parton distributions in heavy-ion collisions? 
In the extreme kinematic limit where gluons
carry a small fraction of the nuclear momentum and become super-dense, it becomes impossible to separate the nucleus into individual 
nucleons. If so, how do we probe this exotic form of partonic matter
in a large nucleus?


\subsubsection{Parton Distributions in Nuclei}

Do quark and gluon degrees of freedom play any role in understanding
the structure of nuclei? In the 1980's the European Muon Collaboration
at CERN demonstrated that the quark momentum distribution 
of a nucleon is significantly altered when it
is placed in a nuclear medium. Recent data from 
DIS indicate that a medium modification also 
occurs in the ratio of the longitudinal to transverse photo-absorption
cross section at low $x$ and $Q^2$. Many models have been proposed
to explain the EMC effect, but no satisfactory
consensus has yet been reached.

Measurements of the nuclear modification of the parton
distributions provide information about the virtual 
particles responsible for nuclear binding. 
If the nucleon-nucleon interaction is mediated by the exchange of 
virtual mesons, it would stand to reason that such exchanges are 
\textit{enhanced} in the nuclear medium. To date, 
however, no such enhancement has been observed.
A nuclear enhancement of valence quarks, sea quarks, 
or gluons would be 
indicative of the relative importance of meson, quark, 
or gluon exchange 
at various distance scales. 
There are as yet no data at $x > 1$ in the scaling region, which
can address the possible existence of super-dense
partonic clusters in the nucleus. 
Also relevant are semi-exclusive experiments, observing
high or low momentum backward nucleons, which can either 
emphasize events originating from superhigh density clusters 
or else pick out events involving an almost unmodified 
neutron target.
Polarization studies of the deuteron
at high energy should determine whether meson exchange or
quark interchange is a dominant process when the two nucleons
are very close together.



\subsubsection{Relevance to High Temperature QCD}

The parton distribution functions in nuclei 
determine the initial conditions for heavy-ion collisions,
which are the only laboratory tool to search for a new
state of matter: the quark-gluon plasma (QGP). The
QGP is a deconfined phase of matter, which is expected 
to occur at very high temperatures and densities, and 
experiments to search for this new phase of matter are underway worldwide.
Significant medium modifications of the gluon distribution 
are certainly expected, but their magnitude is as yet only weakly 
constrained by experiment.
It is of great importance to the 
heavy-ion community that these effects be understood, as they are
an essential ingredient in establishing the observation
of the QGP.

\subsubsection{Partonic Matter under Extreme Conditions}

High energy scattering, with either electromagnetic probes
or protons on nuclear targets, offers new opportunities 
for studying partonic matter under extreme conditions. 
Particularly exciting is the possibility to investigate the 
very-low $x$ region where gluons are dominant.
Measurements of the proton structure function $F_2^p(x)$
have shown that the gluon density rises dramatically as $x$ decreases.
Unitarity considerations indicate that the gluon densities 
must \textit{saturate} at some point, perhaps through the mechanism of 
gluon \textit{recombination}. This new regime
of partonic matter
has not yet been observed and, at the moment, it seems that
we have not yet reached sufficiently low values of $x$. 
However, in heavy nuclei the effects of saturation will be revealed at much larger
values of $x$ than in $ep$ scattering.
New and proposed 
facilities offer for the first time the prospect of reaching the 
gluon-saturation regime and observing this new state of partonic matter.

\section{Spectroscopy}


Spectroscopy is a powerful tool in physics. For example, the color degree of freedom
emerged from detailed baryon spectroscopy and  flavor symmetry was
first seen clearly in hadron spectroscopy. 
The charmonium spectrum solidified our belief in
the existence of quarks and provided substantive evidence for a
linearly confining quark-antiquark potential.  Hadron spectroscopy will
continue to be a key tool in our efforts to understand the long-wavelength
degrees of freedom in Quantum Chromodynamics (QCD).  This section includes an overview of 
experiments and theoretical calculations
for the bound and resonant states of mesons and baryons.
The long-range properties of QCD are central to the
issues of this subfield, bringing into play its full complexity 
\emph{and} a set of rich phenomena in strong interactions.
The properties of QCD and the nature of confinement
are among the outstanding open problems in physics.  
To get a coherent picture, 
contributions from phenomenology, QCD-based models, and
lattice gauge theory (LGT) will be required.
This subject is complementary to the
study of structure functions and is closely linked to 
the hadronic models section.  
Hadronic spectroscopy cannot be explained using standard perturbation
theory.  Nonperturbative field theoretic methods
will be crucial to gain understanding from the data.

Most conventional excited hadrons are regarded as excitations of
the quark degrees of freedom. Theory predicts that the gluonic degrees
of freedom can be excited at moderate energies.  A common language
used for gluonic excitation is that of the flux tube; this
plays a prominent role in many empirical models.  One immediate
manifestation of the possibility of coherent excited glue 
is the presence of \emph{hybrid} states
in the hadron spectrum.  Lattice and model predictions for the mass scale of
these excitations have now converged at, or just below, 2 GeV.
Their quantum numbers, strong decays and production rates in 
electromagnetic processes have been predicted in various model 
and lattice studies.
In particular, the existence of exotic combinations of 
spin, parity and charge-conjugation ($J^{PC}$) quantum numbers
among the hybrid mesons will aid in their identification.
For example, the flux tube model predicts that low lying $1^{-+}$
exotic hybrids have their quarks in a spin triplet. This picture is 
indirectly supported by lattice calculations. If correct, it implies
that exotic hybrids are especially suited to production by photons.

Other new types of hadronic matter are also anticipated. These include
bound states of mesons, states with a $qq\bar q \bar q$ structure,
dibaryons, and $qqqq\bar q$ states. An example of a possible dibaryon
is the $H$ particle, 
whose properties are relevant to the stability of strange matter.
Candidate states exist for all of these nonstandard hadrons.
In a sense, these states interpolate between
hadrons and nuclei and thereby provide an important empirical link between
these regimes. Testing theory and models on these states will thus be a
significant step in developing reliable descriptions of nuclei and nuclear matter.

Many years ago, the discovery of
approximate SU(3) symmetry in the hadron mass spectrum led to 
a breakthrough in establishing the quark model, and is
still a mainstay in particle physics curricula.  As more detailed
information became available, the discussion evolved into 
the issues of interquark forces and details of the baryon wave functions.
It is now clear that nature has given us
an incredible empirical gift, as is evident in the slowly varying 
hadron mass gaps between orbitally-excited 
(e.g. $^3$P$_2$ - $^3$S$_1$)
and spin-excited (e.g. $^3$S$_1$ - $^1$S$_0$) 
mesons as the quark mass evolves from 
heavy to light quarks (or from above the QCD scale to below it).
This surprising
feature motivates the use of nonrelativistic quark models over the full
range of quark masses, despite the fact that the model has no rigorous
justification in light quark systems. 
More recently, advances in computer technology have allowed considerably improved
studies of hadrons using
lattice QCD techniques. One may anticipate that many of the basic aspects
of QCD will be clarified through lattice studies,
and these results may be abstracted by model builders for application
in the regimes of high-mass excitations and scattering, which are not
easily accessible to lattice studies.
In the near term, we expect that progress in hadron
spectroscopy will follow from a synthesis of results from 
lattice gauge theory, empirical quark models, 
and high-statistics 
experiments using partial wave analyses on several final states.

\subsection{Mesons}

The valence content of a meson is understood to be a quark and an antiquark.
This is the basis for 
the constituent quark model (CQM) description of the mesonic structure.  Even more than for baryons, 
this has provided a highly successful empirical
description of the meson mass spectrum and various decays.
This success has the added effect of providing excellent
means to search for exotic hybrid mesons (mesons with
quantum numbers not possible in the CQM) and glueballs (states with a
substantial `pure glue' component).

\subsubsection{Light Mesons}

Much progress has been made in this field recently, both in the 
scalar sector where the lightest glueball is expected 
and in the area of mesons with exotic quantum numbers. 
High quality, very high statistics data at CERN in both $\bar{p}p$
annihilation and $pp$ central production have significantly 
advanced our knowledge of scalar mesons. 
Three states are now known with mass near 1.5 GeV, close to the 
low-lying scalar glueball mass predicted by lattice gauge theory.
The current interpretation is that the
scalar glueball and the meson nonet are strongly mixed in the three
physical resonances (where only two are predicted by the CQM).  
It is encouraging that the two states 
which seem to have the largest glue content are relatively 
narrow. [The $f_{0}(1500)$ has a width of approximately 120\,MeV and
the $f_{0}(1710)$ has a width of about  160\,MeV.]

There have also been reports of states with $J^{PC}=1^{-+}$ 
exotic quantum numbers (which are forbidden to conventional
$q\bar q$ quark model states) in several experiments. 
Experiments at BNL and VES (Serpukhov) have reported an I=1 
$1^{-+}$ exotic resonance with a mass of 1.6 GeV
in three distinct decay modes. There are also recent
reports of the same state in $\bar{p}p$ annihilation at rest. 
A second, more controversial $1^{-+}$ state with a mass of 
1.4 GeV has been reported by two experiments, but 
only in $\eta\pi$ final states.  Although the detailed composition 
these two states is still an open question, especially in view of their
low mass relative to lattice and flux tube predictions for exotic
hybrids, there is no question that (if confirmed) they are beyond
the standard $q\bar q$ quark model.  All of these
observations have been made possible by nearly hermetic experiments
with extremely high statistics combined with 
an excellent understanding of the detectors.  

We have finally reached an era in which we see 
experimental evidence of gluonic excitations. 
In order to understand the physics of these new systems, it
is important establish the spectrum in sufficient detail to see
the pattern of these states. 
The determination of their production and decay characteristics 
will be important to guide future experiments and to
improve our understanding of the physics of gluonic
excitations.  The
clearest place to study decays is in a \emph{clean} arena in which
mixing is minimal.  This means that initially we 
should determine the spectrum of 
states with non-$q\bar{q}$ quantum numbers.  Once these are understood
the analysis can be extended to non-exotic gluonic states that 
may mix significantly with
with conventional hadrons.  Certain clear opportunities for
carrying out this program can be identified.  For example, at present 
there are very little data on the photoproduction of mesons.
The photon
is a very interesting probe for meson production because it
carries a unit of angular momentum into the reaction; 
using a polarized photon beam, exotics and their production mechanism 
can be identified unambiguously.  The widely
held view that hybrid states have excited gluonic flux tubes 
implies that they should have relatively large
photocouplings.  

Glueball and meson calculations on the lattice have advanced 
significantly in the last few years, and these results, in combination 
with recent high statistics data on 
scalar mesons, have changed our interpretation 
of glueball candidates considerably.
The glueball spectrum in Yang-Mills theory 
(pure glue) is now well known;
future calculations in the pure glue sector will address the physical extent
and structure of glueballs.  The inclusion of light-quark
effects is the next challenge in lattice glueball studies, and the 
level of mixing between pure glueball states and
nearby quarkonium states is a crucial issue. Lattice 
results have also been reported for the spectrum of light
exotic mesons; these show that the 
lightest exotic meson has $J^{PC}=1^{-+}$ quantum numbers, and 
a mass near 2 GeV.  Future lattice work will 
improve the statistical accuracy of this result and 
expand the study to other exotic quantum numbers.
Lattice QCD will also address the much more difficult problem of 
meson decays, both for exotic and conventional mesons.

Hybrids are widely expected to be identifiable through their unusual
strong decay amplitudes.  In the flux tube model a hybrid is a state in 
which the flux tube is orbitally excited, and in the usual flux tube breaking
picture of hadron decays this leads to a preference for ``S+P" final states,
such as $f_1\pi$ and $b_1\pi$, over the more familiar ``S+S" modes such as 
$\pi\rho$, $\pi\eta$ and $\pi\eta'$. 
If this prediction is confirmed in exotic hybrids
such as the I=1 $1^{-+}$ $\pi_1$ states, it should be very useful in 
the identification of non-exotic hybrids.  If this selection rule proves 
inaccurate, it will require dramatic revision of the present picture 
of exotics.

Meson form factors continue to be an important testing ground for advanced
models of QCD. New experiments at JLab and Cornell are greatly advancing
our knowledge of the pi meson. Similar efforts are occurring at the AGS,
WASA/CELSIUS (Uppsala) and MAMI (Mainz) for the $\eta$
and $\eta'$ mesons.

\subsubsection{Heavy quark mesons}

The detailed structure of mesons containing $c$ or $b$ quarks 
has traditionally been a high energy physics subject, however 
future opportunities exist for nuclear
physicists.  
The heavy quarkonium and hybrid sectors are especially attractive for the 
study of hadron spectroscopy, since the complications of relativistic 
quark motion and large decay widths, are of
reduced importance. Very interesting results were found in $c\bar c$
mesons above open-charm threshold in the late 1970s and 1980s, such
as a possible $D^*{\bar D}^*$ molecular state. In view of recent theoretical 
and experimental results on exotics it now appears appropriate to 
initiate new experiments 
at a high-statistics $e^+e^-$ ``tau-charm" machine.  
A search for charmed hybrid mesons should be a high priority. 
The lattice implementation of the effective theory which describes 
nonrelativistic QCD has predicted masses of the
$1^{-+}$ exotic heavy-quarkonium hybrids of 4.39(1) GeV for
$c\bar c$ and 10.99(1) GeV for $b\bar b$. A $1^{--}$ hybrid is expected 
to lie close in mass, and given moderate mixing between 
$c\bar c$ and hybrid $c\bar c$, these states should 
appear in $e^+e^-$ experiments in this energy range.  One may also 
search for the non-$1^{--}$
states using a hadronic cascade from higher-mass $c\bar c$ continuum states.
Model calculations
predict that these states will have hadronic decay widths of less than 50 MeV.

Another key idea is to search for glueballs in $J/\psi$ radiative and
hadronic decays.  Previous low-statistics experiments identified glueball
candidates such as the $f_0(1710)$; with high statistics a more definitive
assignment will
be possible through the crucial strong branching fractions of these
states.  Comparison with LGT predictions of glueball couplings to 
meson-meson final states should allow discrimination of glueballs 
from other types of states, or the identification of the glue component 
of strongly mixed states.  

A third interesting direction is the study of photon-photon collisions at high intensity
$e^+ e^-$ experiments such as CLEO and BABAR.
Previous results 
have been very useful in flagging non-CQM properties of unusual states such 
as the $a_0(980)$, $f_0(980)$ and $f_0(1500)$, which have anomalously 
small $\gamma\gamma$ widths compared to known $q\bar q$ states. 
Theoretical predictions for $\gamma\gamma$ widths have been tested for
about 10 light 1S and 1P $q\bar q$ and $c\bar c$ states,
but the very limited statistics to date have precluded detailed studies
of the scalar $f_0$ states, which would 
clarify the nature of glueball candidates.  Finally,
knowledge of the $\eta'$ form factor 
derived from radiative decay measurements may help solve the puzzle of its
mass generation.

Heavy-light meson systems ($D$,$D^*$,$D_s$,$B$,...) 
have been studied in detail using heavy quark effective theory (HQET), 
especially their weak transition amplitudes, 
which are described by the Isgur-Wise function.  Many 
other interesting predictions of HQET
motivate identification of the higher mass, strongly 
unstable heavy-light states.  

CP studies at high energy machines are high priority experiments that
depend on detailed understanding of various strong decays.
Recently, complications in the determination of CP phases due to strong 
final state interaction (FSI)
effects have been realized.  Studies of
$D$ and $D_s$ decays have confirmed experimentally that these FSI phases are 
important. A better understanding of strong interaction effects in the
CP-relevant channels will be required. Similarly, determination of
Cabibbo-suppressed CKM matrix elements and $D\bar D$ mixing parameters
will require an understanding of strong interaction effects among the light
hadron decay products of heavy-light mesons.

\subsection{Baryons}

The most basic elements of baryon spectroscopy are the ground state
properties of the proton: mass, spin, magnetic moment, charge radius.  
The main goals of modern experiments are
the full determination of the spectrum of excited states,
identification of possible new symmetries in the spectrum, and illuminating
the microscopic structure of states that are nominally
built of three valence quarks.  As mentioned above, the 
establishment of SU(3) symmetry was a key result in particle
physics at the beginning of baryon structure studies.  Advanced
experimental capabilities and the ability to solve models with
close approximation to QCD now allow far deeper understanding.
The present status will be discussed in this subsection.

Many fundamental issues in baryon spectroscopy are still not 
well understood.  This is largely due to the lack of data
beyond the early HEP experiments of a few decades ago.  
The very limited knowledge of states beyond the lowest S and P wave 
supermultiplets provides very weak constraints on 
models.
The possibility of new, as yet unappreciated, symmetries could be 
addressed with better data.  For example, there may be a parity doubling
in the spectrum of baryons, which would be observable 
in different flavor sectors.  If parity doubling is a real effect, 
this implies that
the usually anomalous axial $U(1)$ symmetry has been 
restored.  An investigation of this possibility through a search for 
additional states would help clarify this issue. 
There is also an outstanding controversy over whether all three
quarks in a baryon can be excited, or whether a quark-diquark
picture is more appropriate.  One can definitively distinguish 
between symmetric quark models and strict quark-diquark models through
the discovery of a comparatively small number of new positive parity 
excited states, which are
predicted by the symmetric $qqq$ models but are absent from the
$q(qq)$ diquark models.
If these states exist, they are expected to appear strongly in
certain novel final states such as $N\eta$ and $N\omega$.
This issue should be resolved by a careful analysis of data obtained from
a variety of initial and final states.


The other major direction is to use spectroscopic information
to learn about the underlying forces that act on quarks in baryons.
The mass spectrum displays the ordering of states by
spin, parity, and flavor.  This can be thought of as empirical
splittings that provide information about the effective degrees
of freedom and has already provided the basis
for many empirical models.  The decay branching fractions of excited
baryons to various asymptotic states and the corresponding angular
distributions provide more detailed filters for models.  A
key tool for spectroscopy is the photocoupling amplitude, $\gamma N \rightarrow N^*$.  Unlike its analog in 
atomic spectroscopy, this is an excitation amplitude. 
It can then be measured as a function of photon 4-momentum ($Q^2$)
and provide additional structure information. 

In recent years,
several labs in the US (BNL, JLab) and Europe 
(Mainz, Bonn) have initiated vigorous programs in baryon spectroscopy.
The use of modern detectors with large acceptance -- effectively electronic
bubble chambers -- 
and high statistics capabilities will jump start 
important advances in our knowledge of baryon
spectroscopy.  The important new detectors
are just starting to publish data, so although the
picture is far from complete, the quality of the new
results has been demonstrated.

The early JLab results for baryon spectroscopy 
involve much better statistics and kinematic coverage than
all previous experiments combined. 
Additional reactions are being measured for the first time.
Empirical analyses have determined initial
results for the photocoupling amplitudes over a broad range of $Q^2$.
Polarization of beam and target
are expected to play a key role in disentangling the spectrum.
At BNL, the Crystal Ball collaboration is measuring the $\Lambda$ and $\Sigma$
hyperon spectra at low energy, as well as $N^*$ and $\Delta^*$ properties.
The accuracy of the results is much higher than that of any previous experiment.
A wealth of new data on $\pi p$ and $K p$ reactions is being produced which will
greatly improve our understanding of the light  baryons, and especially of the poorly
known $\Sigma$ states.
With a careful partial wave analysis effort (spearheaded, for example,  by
the recently formed Baryon Resonance Analysis Group)
a detailed description
of the baryon spectrum for masses below about 2.2 GeV is sought. 
A unified analysis of hadronic and electromagnetic reactions is required to 
unambiguously extract underlying physics.
Hadronic beams will also be central in the experimental resolution of
fundamental issues in QCD. For example, several processes are sensitive
to quark mass ratios or differences. These include isospin forbidden pion
production in deuteron-deuteron collisions and the decay $\eta \rightarrow 3\pi$.

Calculations are largely made with models that have quite different
assumptions -- constituent quark models, continuum models,  and lattice gauge
theories (see section 5).  Each has value in elucidating features of the underlying
dynamics.  Lattice gauge theory  solves QCD with a very small set of assumptions
while constituent quark models  assume the existence of massive quarks as the most
significant degree of freedom.
Although the empirical constituent quark models do not have a clear 
derivation from QCD, they nonetheless appear to 
incorporate much of the relevant physics of strong QCD with a
small number of parameters.
Given these parameters, the model predicts the spectrum of a wide 
range of heavy and light mesons and baryons with sufficient 
accuracy to make it very useful for the interpretation of experimental
data on resonances.  Perhaps the most significant success for baryons
is the qualitatively correct description of almost all (about 50)
$\gamma N \rightarrow N^*$ photocouplings.

The JLab/MIT lattice group has recently begun a program of LGT
$N^{*}$ calculations, with a detailed study of several of the 
lowest lying baryon states as a principal objective.
Calculations of the masses of a few of the lowest lying 
states with angular momentum up to $5/2$ should be completed
in the near future. Work is also in progress on the 
photocouplings of these states.  These studies will 
confront the emerging high quality data from the new series of
baryon experiments.  They will also provide information about 
more qualitative aspects of baryon models, such as the 
appropriate fundamental degrees of freedom in the baryons, for 
example $qqq$ versus $(qq)q$.  Any disagreement between 
lattice QCD and experiment would be a striking and perhaps far-reaching
discovery.

There are important roles for both constituent quark models and lattice
gauge theory
in describing the many phenomena seen in meson and
baryon spectroscopy. At present, neither the CQM nor LGT
satisfy the needs of the field for accurate, well-founded 
descriptions of the 
spectrum of low-lying and highly-excited
states and their production and decay properties.  
First-principles lattice studies (in the quenched approximation) 
will soon yield
much of the low-lying meson and baryon mass spectrum, but QCD-based models
will be needed for guidance with respect to production
and decay amplitudes.



\section{The Chiral Structure of Matter, Form Factors, and Few Body Nuclei}

\subsection{The Chiral Structure Of Matter}

The study of the chiral structure of matter is an active and
fundamental field. The relevant phenomena are the properties of the
Goldstone bosons as probed by their interactions and production amplitudes.
These are rigorously linked to QCD by
an effective (low energy) field theory, chiral perturbation theory (ChPT).

The chiral limit of QCD refers to the limit in which the bare light quark masses
are zero. In this limit, QCD exhibits a `chiral' symmetry which is not manifest in
nature (and is therefore `hidden' or `broken'). A fundamental theorem then
implies that massless (Goldstone) bosons must exist in the excitation spectrum of the theory.
In the case  of massless $u$, $d$, and $s$ quarks, these are the pions, eta, and kaons.
Non-zero light quark masses explicitly break the
chiral symmetry of the Lagrangian with the result that the pion, eta, and
kaon have finite masses. 
In the chiral limit, these Goldstone bosons do not interact with hadrons at very low
energies, thus the small low energy interactions that are measured probe finite
quark mass effects in QCD.
In particular, the electromagnetic
production amplitudes and the internal properties (e.g. radii,
polarizabilities, decay widths will serve as fundamental tests of the chiral
structure of matter. These measurements represent timely physics issues and a
technical challenge for experimental physics.


The long standing prediction of Weinberg that the mass difference of the
up and down quarks leads to isospin breaking in $\pi N$ scattering is of 
special interest in this field. The accuracy of the completed experiments 
and of the model extractions from the deuteron pionic atom, does not yet 
permit a test of this fundamental prediction. An interesting possibility 
is the use of the pion photoproduction reaction with polarized targets 
to measure the isospin breaking predictions of low energy $\pi^0 N$
scattering, which is related to the isospin breaking quantity $\frac{m_d
-m_u}{m_d+m_u}$ discussed in the introduction to this document.

Not all chiral predictions have been properly tested. The
experimental magnitude of the  $\pi N\ \Sigma$ term is still
uncertain. This is a
fundamental quantity which gives a measure of the strange quark
contribution to the nucleon mass. Important, precise experiments in low
energy pion-nucleon scattering and charge
exchange are presently being performed and are also in the planning stage.
Another unsolved problem is a contradictory experimental situation for the
pion
polarizabilities. 
Experiments on $\eta$ and $K$ production and
scattering  are in their infancy. They require both the high quality of
existing beams and experiments cleverly designed to reduce the resonance 
contributions. Measurements of the
$\eta N$ and $KN$ interactions would provide important tests of the quasi
Goldstone boson nature of these heavier pseudoscalar mesons. 

Some fundamental nucleon properties (for example,  electromagnetic
polarizabilities) diverge in the chiral limit, indicating that they are pion 
dominated. Measuring these
with
real and virtual photons allows one to make a detailed map of their spatial
distributions. The study of the non-spherical amplitudes in the nucleon
and
$\Delta$ wave functions also reflect significant non-spherical pion field
contributions, as expected from Goldstone's theorem.

A profound example of symmetry breaking in QCD is the axial
anomaly. The classical U(1) symmetry of the QCD Lagrangian is absent in
the
quantum theory presumably due to quantum fluctuations of the quark and
gluon
fields. Physical consequences are the non-zero mass of the $\eta$
meson and the 2 photon decays of the pseudoscalar mesons.  There is an
absolute prediction of the $\pi^0 \rightarrow \gamma \gamma$ decay 
rate with only one parameter, $N_c$, the number of colors in QCD. 
At present the accuracy of the
experiments is approximately 15\%. An effort to reduce this by an order of
magnitude is in progress. There are also plans being made to measure the
$\eta\rightarrow\gamma \gamma$ to and
$\eta' \rightarrow \gamma \gamma$ decay rates. These involve the axial
anomaly and also the mixing between $\pi^0$, $\eta$, and
$\eta'$ mesons (which vanishes in the chiral limit). The present
experimental accuracy of the $\eta$ and $\eta'$ two photon decay rate  is
approximately 15\%. It appears feasible to reduce these errors by about a
factor of 5. This would
 significantly improve the determination of the mixing matrix.
Other reactions for which the axial anomaly is the dominant
mechanism such as $\gamma \pi \rightarrow \pi  \pi$ are also
being studied.

\subsection{Nucleon Electromagnetic Form Factors}

The electromagnetic form factors of the nucleon have been of
longstanding interest in nuclear and particle physics.
Form factors describe
the distribution of charge and magnetization within
nucleons and allow sensitive tests of nucleon models based on
Quantum Chromodynamics or lattice QCD calculations.
They also are important input for calculations
of processes involving the electromagnetic interaction with complex nuclei.
Precise data on the nucleon
electromagnetic form factors are essential for the analysis of parity
violation experiments, designed to probe the strangeness content of the
nucleon. The nucleon electromagnetic form factors are closely related
to the newly discovered generalized parton distributions.
Thus, the study of the nucleon electromagnetic form factors
advances our knowledge of
nucleon structure and provides a basis for the understanding of more
complex strongly interacting matter in terms of quark and gluon
degrees of freedom.

The proton electric ($G^p_E$) and magnetic ($G^p_M$) form factors 
have been studied extensively in the
past from unpolarized electron-proton elastic scattering
using the Rosenbluth separation technique.
The maturation of polarization methods has revolutionized our ability to 
study electromagnetic structure.
For example, the standard dipole parameterization
seemed to describe the $Q^2$ 
dependence of both proton form factors well at low momentum transfer. 
However, new data from a polarization transfer experiment at JLab directly
measures the ratio ${\frac{\mu G^p_E}{G^p_M}}$.
Strong disagreement with the dipole form factors at moderate 
momentum transfer is found, necessitating a reassessment of the longstanding picture of 
the reaction. 



Until recently, most data on $G_M^n$ had been deduced from
elastic and quasielastic electron-deuteron scattering experiments. 
For inclusive measurements, this procedure
requires the subtraction of a large proton contribution and suffers
from large theoretical uncertainties. 
The sensitivity to nuclear structure can be greatly
reduced by measuring the cross section ratio $d(e,e'n)/d(e,e'p)$ at
quasielastic kinematics.  
While the precision of recent experiments at Mainz and Bonn is excellent, 
their results are not fully consistent. 
An alternative approach for precision measurements of $G_M^n$ uses
the inclusive quasielastic $^3\vec{\rm He}(\vec{e},e')$ process. 
By using polarization observables, these measurements are subject 
to different systematics
than the unpolarized deuterium experiments.  Experiments at various labs are
in progress.

The intriguing result on ${\frac{\mu G^p_E}{G^p_M}}$ at high $Q^2$ 
elicited great interest in this 
subject. An extension to $Q^2$ = 5.6 GeV$^2$ is currently 
in progress at JLab. A precision measurement 
of the proton RMS charge radius is planned at Bates 
with BLAST to take advantage of the possibility of 
precision lattice QCD calculations.
The planned measurement will improve the precision of $r_p$ 
by a factor of three compared with the single most precise 
measurement from electron scattering experiments.
This will be sufficient
to allow high precision tests of QED from 
hydrogen Lamb shift measurements and 
to provide reliable tests of
lattice QCD calculations.

Unlike the proton electromagnetic form factors, 
data on the neutron form factors
are of inferior quality due to the lack of free neutron targets.
However, recent experiments have demonstrated that $G^n_E$ can be
determined with much better precision using polarization degrees of 
freedom.
Polarization experiments are in progress and planned at JLab on
$G^n_E$ for high $Q^2$, and for low $Q^2$ at Mainz, Bates, NIKHEF, and Bonn.
This will allow the most
precise search for the predicted modification of the neutron pion cloud
in the nuclear medium. With future new precision data on $G^n_E$ from 
Jefferson Lab, Bates and Mainz, our knowledge of the neutron charge 
distribution will be improved to a level comparable to that of the proton.
A possible future energy upgrade of CEBAF to 12 GeV at JLab would
allow extension of nucleon electromagnetic form factor 
measurements to much higher $Q^2$ values.
  
\subsection{Few Body Nuclei}

Form factors in elastic electron scattering have been essential in
the investigation of nucleon and nuclear structure.  
At very low momentum 
transfer, the charge and magnetic radii of the nucleon and nuclei can be
determined from the form factors.  
Form factors not only inform us
of bulk properties such as charge and magnetic radii, but also provide the shape of 
the probability distributions.
At low momentum transfer, 
the quarks in a nucleus congregate into nucleons and the 
traditional meson-nucleon picture of the nucleus describes the form factors
well.  However, at high momentum transfer on the simplest nucleus, the deuteron,
it appears that the newest deuteron form factor data are consistent with the
perturbative quark counting rule picture as well as the meson-nucleon picture.
In order to resolve these two pictures, it is essential to extend
form factor measurements of the light nuclei to the highest possible
values of momentum transfer, where sensitivity to quark degrees of freedom are
expected to 
be enhanced.

Another avenue for investigating the role of quarks in nuclei is
photodisintegration of the deuteron.  In general, the momentum transfer
given to the constituents in photodisintegration can be substantially 
larger than 
that in elastic electron scattering because of the large momentum 
mismatch between the incoming photon and the constituents of the nucleus.
Thus, one might expect to see the effects of QCD in photodisintegration
with beam energies of a few GeV.  Indeed, it appears that recent JLab data 
for high 
energy exclusive break up of the deuteron data are consistent with the 
constituent counting
rules while meson exchange models have failed to explain these
high energy data.  Recent polarization data in deuteron photodisintegration
also show a very interesting and seemingly simple behavior.  The
induced proton polarization in deuteron photodisintegration vanishes
at high energies.  This is unexpected from a 
meson-nucleon picture of deuteron photodisintegration because of the
presence of known excited states of the nucleon.  These effects should
be explored at high energies and at a complete set of reaction angles to
determine whether we are actually seeing our first glimpse of 
the transition region between the
nucleon-meson picture and a quark-gluon picture of a nuclear reaction.


\section{Models of the Quark Structure of Matter}

A wealth of experimental data is being collected in hadronic spectroscopy
and deeply inelastic scattering experiments. Although these data are all correlated
by the QCD Lagrangian, it is generally agreed that the dynamics of QCD make it
difficult to understand the data from first principles. However, it is also
believed that the majority of these data may be understood in terms of appropriate
effective degrees of freedom. It is the job of models to determine these degrees of 
freedom, to understand their dynamics, and to employ this understanding to reliably
examine new phenomena.
These are necessarily
scale-dependent questions; this scale dependence can be broken into
three regions: structure, at scales $Q\ll 1$ GeV;
substructure, at $Q\simeq 1$ GeV; and the partonic region, at scales
$Q\gg 1$ GeV.  
An important requirement of all such models is a firm  connection to the vacuum 
structure of QCD.

\subsection{Structure}

A natural language for the description of the `structure' 
  of hadrons is the {\it constituent quark} model, where constituent-quark
effective degrees of freedom interact  via potentials, {\it flux
  tubes}, or are confined to a bag. 

The notion of
the constituent quark comes from phenomenology; for this reason
its definition is necessarily model dependent. Nevertheless, the
success of this phenomenology indicates that some of the properties of
constituent quarks, such as their effective masses and sizes, may be
derivable from QCD.
This is where the connection of
models to experiment is the closest.  Relating model predictions to
data involves the calculation of observables through models of
reaction dynamics, the estimation of final-state interactions and the
effects of open decay channels on hadron masses and properties, and
the development of tools for partial-wave analysis. This study should
ultimately evolve into a description of low-energy hadron-hadron
interactions and nuclear forces.

An important aspect of the analysis of hadron structure involves the
study of pair creation, which has an impact on the strong decays of
hadrons, and fragmentation.  
At present, a microscopic basis for models of strong decays is lacking.
The development of these models has
been driven largely by meson strong decays, with some evidence from
baryon decays. New information about the nature of these decays can be
found from examining excited baryon decays and those of {\it exotic mesons},
and the nature of the strong production process can be accessed
theoretically using lattice QCD. Exotic mesons and {\it hybrid baryons}
should be studied in more detail on the lattice, which allows an efficient
description of excited gluonic degrees of freedom and the mapping of 
the hadronic wave functions. There is a class of models which
employs pions as effective degrees of freedom. Such models emphasize
chiral symmetry and have typically been used to study $N$ and $\Delta$
static properties. The extension of such models to describe excited
states with an accuracy comparable to the constituent quark model
remains an important challenge.

\subsection{Substructure}

The primary concern of this study is identification of the
effective degrees of freedom and their interactions in terms of the
fundamental fields of QCD.  First principles analysis of QCD reveals
several phenomena which can be used to connect to the constituents
responsible for hadron structure.  These include flux tubes, the
constituent quarks themselves, and also topological field
configurations such as {\it instantons}, {\it monopoles} and {\it
  center vortices}.

Constituent quarks appear as a consequence of dynamical chiral
symmetry breaking, as visualized by models like that of Nambu and  Jona-Lasinio.
The instanton liquid model provides a description of the constituent
quarks along with a mid-range spin-dependent interquark force
 as well as OZI violating amplitudes. It naturally incorporates $U(1)$
symmetry breaking, and is capable of connecting to the partonic
degrees of freedom. Progress is needed to
verify the importance of instanton-like configurations for low-energy
quark-antiquark and diquark interactions. These issues can be
addressed using methods based on analysis of the near-zero-mode
eigenvalues of the Dirac operator on the lattice; which will also lead to
an understanding of chiral symmetry breaking.

Flux tube like structures of the gluonic fields are believed to be 
responsible for confinement. Flux tubes provide a natural basis for the study of
hadrons containing excited glue. The main approaches which lead to
flux tubes are the strong-coupling expansion on the lattice and an
effective description in terms of a dual superconductor, which may be
due to condensation of monopoles or central vortices. Further studies
of the internal structure of flux tubes on the lattice and in models are needed. 

Given the utility of the large $N_c$ limit of QCD as a tool for
organizing the magnitudes of effects in low-energy QCD, it is important to
test conclusions based on the large $N_c$ limit on the lattice. For
example, it would be useful to study the behavior of the $\eta^\prime$
mass for $N_c>3$ on the lattice. Similarly, chiral
perturbation theory should be connected to models and lattice QCD by
calculations of chiral expansion coefficients. The method of QCD sum
rules has proven a useful tool in revealing a connection between the QCD
vacuum structure and hadron phenomenology.  Finally, effective field
theories (EFT) have enjoyed a resurgence of interest over the past several
years. These carry the promise of providing a rigorous methodology for
examining QCD in various limiting regimes (heavy quark, low energy, etc).
In addition, the confrontation of theoretical calculations with experimental
measurements in any of these regimes is most efficiently carried out by
comparing the calculated and measured  EFT constants rather than by comparing
prediction and experiment on a case by case basis.

Continuum field-theoretic models of QCD in a fixed gauge, such as
light-cone or Coulomb-gauge, and Schwinger-Dyson Bethe-Salpeter models 
provide an important approach towards understanding hadron
substructure.  They address all of the relevant features of QCD such
as the structure of the vacuum, chiral-symmetry breaking, confinement
and strong decays, as well as hadronic interactions and the connection
to the parton model. 
Indeed important progress has recently been made in understanding QCD and 
hadron physics as a
problem in continuum quantum field theory.  For example, while the fundamental question of
the connection between QCD and the Hilbert space of observable states remains
unanswered, Hamiltonian light-front methods have made promising steps toward
providing a direct connection between QCD and constituent quarks. They are also
making progress in 
calculations of the $x$ dependence of light-front wave functions.
Light-cone wavefunctions (LCWF) provide a fundamental
frame-independent description of hadrons in terms of their quark and gluon
degrees of freedom at the amplitude level. 
Furthermore, the generalized form factors measured in deeply virtual Compton
scattering are
given by overlaps of light-cone wavefunctions. 
The light-cone wavefunction representation also provides a basis for
describing nuclei in terms of their meson and nucleon
degrees of freedom, thus providing a rigorous basis for relativistic
nuclear physics. 
The most challenging problem confronting light-cone theory is the
calculation of hadron LCWFs from first principles in 3+1 dimensional QCD.
Progress is being made with
light-front Hamiltonian 
quantization methods such as discretized light cone quantization and the transverse
lattice. Finally, the role of zero modes
in understanding the condensate and chiral symmetry breaking in QCD on the light front also 
remains to be clarified.

Similarly, Dyson-Schwinger equation studies have made important advances, for example
they provide a microscopic
understanding of the dual nature of the pion as a Goldstone boson and a $q
\bar q$ bound state.  Analyses and modeling of continuum Coulomb gauge QCD
can also address the dual nature of the pion and  have
supported the glueball spectrum obtained in lattice QCD simulations. This
is one example of the potential for positive feedback between lattice
simulations and continuum studies.  Another is the recent lattice computation
of dressed-quark and gluon propagators, which are important elements of
continuum phenomenology.  Since such theory and phenomenology can rapidly
adapt to an evolving experimental environment they must continue to actively 
assist these programs. Important open questions for all continuum methods  are
the related issues of nonperturbative computational methods and nonperturbative
renormalization. The technologies of effective field theory and lattice renormalization
are greatly aiding in resolving the latter problem.

Lattice studies at this important scale should be extended to provide 
correlation functions of different operators. Using high accuracy
data, such as vector and axial-vector amplitudes from $\tau$ decays, 
 one can accurately calibrate lattice calculations.

\subsection{Partonic region}

This region of QCD structure corresponds to the parton
model of deeply inelastic scattering (DIS), at large momentum transfers
$Q \gg 1$ GeV.  The models described above should also be explored in
this regime.  One important issue is to understand the connection of
inclusive observables, such as structure functions, to exclusive
observables such as form factors.  A related issue is to understand
how partons are distributed in multiplicity.  An understanding of how
partons are distributed in the transverse plane is also required. 
These issues can be experimentally addressed by study of diffraction,
multiple-parton collisions and semi-inclusive production.

An important synthesis needs to be made between the intermediate-momentum
transfer (1 GeV) behavior of the parton distributions, and models of the
structure and sub-structure of hadrons. One way to do this is a higher
twist analysis similar to what has been done for correlators in the
framework of QCD sum rules. Bloom-Gilman duality provides
another important connection of this behavior to an effective
description in terms of hadrons.  For example at low $Q$ and in the
large $N_c$ limit the intermediate inelastic states in a calculation
of DIS in the rest frame of the proton are sharp baryon resonances, while at
high $Q$ the large number of high-mass resonances accessible forms a
complete set that gives Bjorken scaling through a  change to a
partonic basis. These can be tested
with detailed experimental information on the low-$Q$ limit and
evolution of structure functions. 

A fundamental aspect of a hadron is its
distribution amplitude $\phi_H(x,Q)$, which controls large momentum
transfer exclusive reactions.  
Factorization theorems have recently been proven, which also allow one
to rigorously compute types of exclusive $B$ decays in terms of the
distribution amplitudes of the final state hadrons.  
The fact that LCWFs are process-independent provides a profound connection
between amplitudes that describe exclusive processes such as
elastic form factors, two-photon reactions, and heavy hadron decays.

Other important quantities that should be better addressed by all
relevant models and lattice calculations are the primordial structure
functions used as input to QCD evolution. In particular an
explanation for the observed sea-quark isospin and spin structure
should be sought. New measurements, such as the spin-flavor antiquark
asymmetry $\Delta{\bar d}-\Delta{\bar u}$, should be performed.

\section{Tools}

Although some of the challenging questions laid out in the
previous sections are decades old, we are at a threshold for 
making significant progress
in resolving these puzzles. This is true, to a large extent, because of the
unprecedented experimental and theoretical tools which are now at our disposal.
These new opportunities have been made possible by recent technological advances. 
For example, lattice field
theory has developed into a powerful and essential tool to understand and
solve QCD. A new generation of accelerators and detectors make
possible experiments with unprecedented precision and kinematic
range. Taking full advantage of this emerging technology will have
decisive impact on hadronic physics.

\subsection{Experiment}


An essential foundation for progress in hadronic physics is the
aggressive exploitation of present facilities and development of new
ones, with a clear focus on experiments that provide genuine insight
into the inner workings of QCD. 
In the near term, it must be a high priority to fully exploit existing
modern facilities. At JLab, a 12 GeV energy upgrade will
open new windows on hadronic physics. At BNL completion of the 
detectors will make possible full exploitation of the unique hadron 
beams at RHIC. A new initiative in lattice QCD at the scale of 
10 Tflops is required to exploit new advances in lattice field theory.

In addition to the present dedicated facilities supported by the Nuclear
Physics program, significant opportunities exist to use lepton and
hadron beams at other accelerator facilities, for example at Fermilab,
the BNL-AGS, and CLEO. It is important for
our field to aggressively utilize capabilities of these beams to 
address key issues in hadronic physics.
The breadth of the effort required to give an accurate picture of
the meson and exotic spectrum makes results from a variety of labs
valuable. 

In the longer term, a high luminosity electron ion collider would be
a powerful new microscope for the examination of  hadronic structure. To develop the 
optimal capabilities in a timely way, research and development of
accelerator and detector technology will be necessary.

\subsection{Theory}

One of the principal reasons this is a
propitious time for fundamental progress in hadronic physics is that the
tools of lattice field theory and the availability of Terascale computers
now make definitive calculations of hadron observables possible.
Algorithms that incorporate chiral symmetry exactly on the lattice,
chiral perturbation theory to extrapolate reliably from the
masses at which lattice calculations are performed to the masses relevant
to the physical pion mass, and Terascale computational resources provide 
an unprecedented opportunity for controlled solutions, which
 will have decisive impact on our understanding of QCD. 

Similarly, advances are being made in continuum model building, effective
field theory, and the theory and application of parton distributions.
As a result, increased support of
theory is warranted. This could include strengthening university and laboratory 
based research,
the creation of a national postdoctoral fellowship program in hadronic physics,
increased support of bridge positions, 
laboratory visitor programs, and the support of summer schools for undergraduate
and graduate students.

\vskip .5 true in
\noindent
{\bf Acknowledgements}

The workshop participants would like to acknowledge the financial
support of the Thomas Jefferson National Accelerator
Facility, Argonne National Laboratory, MIT/Bates, Los Alamos 
National Laboratory, and Brookhaven National Laboratory.

\vskip .5 true in

\centerline{\bf Contributors}

\begin{tabbing}
Shaliesh Chandrasekharan  \=Old Dominion Univ/Jefferson Lab   \kill
Peter Barnes    \>Los Alamos National Lab \\
Ted Barnes    \>Oak Ridge National Lab \\
Aron Bernstein \> MIT\\
James Bjorken   \>SLAC    \\
Stan Brodsky    \>SLAC \\
Matthias Burkardt \>New Mexico State University \\
Simon Capstick   \>Florida State University \\
Lawrence Cardman   \>Jefferson Lab   \\
Carl Carlson    \>College of William \& Mary \\
Shailesh Chandrasekharan  \>Duke University\\
Frank Close    \>Rutherford Laboratory \\
Tom Cohen    \>University of Maryland \\
Kees de Jager    \>Jefferson Lab  \\
John Domingo    \>Jefferson Lab \\
Steve Dytman    \>University of Pittsburgh\\
Alex Dzierba    \>Indiana University  \\
Robert Edwards   \>Jefferson Lab \\
Rolf Ent    \>Jefferson Lab/Hampton University\\
Liping Gan \> Jefferson Lab  \\
Ashot Gasparian \> Jefferson Lab   \\
Haiyan Gau \> MIT \\
Barry Holstein \> University of Massachusetts \\
Roy Holt    \>Argonne National Lab \\
Donald Isenhower   \>Abilene Christian University\\
Nathan Isgur    \>Jefferson Lab  \\
Robert Jaffe    \>MIT   \\
Sabine Jeschonnek   \>Jefferson Lab \\
Xiangdong Ji    \>University of Maryland \\
Ed Kinney    \>University of Colorado \\
Leonard Kisslinger \>Carnegie Mellon University \\
Frank Lee    \>George Washington University \\
Mike Leitch    \>Los Alamos National Lab  \\
Naomi Makins    \>UIUC     \\
Mark Manley \> Kent State University \\
John McClelland   \>Los Alamos National Lab \\
Larry McLerran   \>Brookhaven National Lab \\
Wally Melnitchouk   \>University of Adelaide  \\
Mac Mestayer    \>Jefferson Lab   \\
Curtis Meyer    \>Carnegie Mellon University \\
Chris Michael    \>University of Liverpool\\
Richard Milner   \>MIT   \\
Colin Morningstar   \>Carnegie Mellon University\\
Joel Moss    \>Los Alamos National Lab \\
Ben Nefkens    \>UCLA    \\
John Negele    \>MIT  \\
Jen-Chieh Peng  \> Los Alamos National Lab  \\
Anatoly Radyushkin   \>Old Dominion Univ/Jefferson Lab \\
David Richards   \>Old Dominion Univ/Jefferson Lab \\
Craig Roberts \>Argonne National Lab\\
Edward Shuryak   \>SUNY-Stonybrook   \\
Hal Spinka \> Argonne National Lab \\
Paul Stoler \> Rensselaer Polytechnic Institute \\
Mark Strikman   \>Penn State University   \\
Eric Swanson    \>University of Pittsburgh  \\
Adam Szczepaniak   \>Indiana University  \\
Hank Thacker    \>University of Virginia \\
Frank Wilczek    \>MIT 
\end{tabbing}

\centerline{\bf Editors}

\begin{tabbing}
Shaliesh Chandrasekharan  \=Old Dominion Univ/Jefferson Lab   \kill
Simon Capstick \>Florida State University \\
Steven Dytman \>University of Pittsburgh \\
Roy Holt    \>Argonne National Lab \\
Xiangdong Ji    \>University of Maryland \\
John Negele    \>MIT  \\
Eric Swanson    \>University of Pittsburgh 
\end{tabbing}

\end{document}